# Tiling for Performance Tuning on Different Models of GPUs

Chang Xu
Department of Information Engineering
Zhejiang Business Technology Institute
Ningbo, China
colin.xu1982@gmail.com

Steven R. Kirk, Samantha Jenkins
Department of Economics and IT
University West
Trollhattan, Sweden
(steven.kirk, samantha.jenkins)@hv.se

*Abstract*—**The strategy of using CUDA-compatible GPUs as a parallel computation solution to improve the performance of programs has been more and more widely approved during the last two years since the CUDA platform was released. Its benefit extends from the graphic domain to many other computationally intensive domains. Tiling, as the most general and important technique, is widely used for optimization in CUDA programs. New models of GPUs with better compute capabilities have, however, been released, new versions of CUDA SDKs were also released. These updated compute capabilities must to be considered when optimizing using the tiling technique. In this paper, we implement image interpolation algorithms as a test case to discuss how different tiling strategies affect the program's performance. We especially focus on how the different models of GPUs affect the tiling's effectiveness by executing the same program on two different models of GPUs equipped testing platforms. The results demonstrate that an optimized tiling strategy on one GPU model is not always a good solution when execute on other GPU models, especially when some external conditions were changed.**

## I. INTRODUCTION

In 2007, NVIDIA released its programming architecture named CUDA (Compute Unified Device Architecture) [1] for developing parallel programs with their programmable GPUs (Graphics Processing Unit). CUDA is a general purpose parallel computing architecture that leverages the parallel compute engine in GPUs to solve many complex computational problems in a fraction of the time required on a CPU [1]. Before the release of CUDA, attracted by the powerful parallel compute ability of GPUs, many developers and communities proposed solutions such as some of the traditional GPGPU (General Purpose Graphical Processing Unit) [2] and some other traditional GPU programming models [3] to use GPUs to perform high performance parallel computing. Few of these approaches, however, can offer acceptable effectiveness and efficiency [3] [4]. The appearance of CUDA and CUDA compatible GPUs motivated a revolution in this field. It offers an effective and efficient platform for programmers to develop many kinds of high performance parallel applications. At the same time GPUs stepped out of their traditional graphic field and into other fields requiring high-performance computation, such as hydromechanics, medicine, and simulation and so on [4]. In order to address these new demanding markets, NVIDIA also released a new series of evolved GPU specified for high performance computation—Tesla[1]. NVIDIA has also recently announced their plans for a further improved GPU architecture, codenamed 'Fermi'.

Coming together with the benefits of CUDA is its flexibility in programming. In some sense, the flexibility is a double-edged sword: it offers many optimizations for programmers to tune code to get better performance but on the other hand it brings challenges for programmers to develop. Many optimization techniques are mentioned in the literature [4] [5] [6] [7] and also in the NVIDIA CUDA Programming Guide [8]. Categories of optimizations can be summarized as tiling, using shared memory, unrolling and prefetching. Among these, tiling is the most basic but also most important technique and it is always the decisive factor that affecting programs' performance [4] [6].

As well as the release of new models of GPUs, different versions of compute capabilities implementing different features were delivered. Table I lists several different features those we should primarily consider when developing CUDA program. These features will affect the design of tiling when tuning a program's performance. So the problem is raised as whether a tiling dimension which can provide best performance for a program wile executed on different GPU models.

TABLE I. COMPUTE CAPABILITY OF GTX260 AND GEFORCE 8800

| Features | GTX 260 | GeForce 8800 GTS |
|---|---|---|
| number of register per SM | 16384 | 8192 |
| active warps per SM | 32 | 24 |
| active threads per SM | 1024 | 768 |
| total SP | 192 | 96 |
| number of SM | 24 | 12 |
| global memory | 1G | 320M |

[1]Tesla is NVIDIA's first dedicated General Purpose GPU for high performance

## II. BACKGROUND

### A. *Programming Model of CUDA*

In CUDA terminology, the CPU is called the host and the system memory is called host memory [9]. Correspondingly the GPU is called the device and the global memory on the graphic card is called device memory. The codes in CUDA are divided into two main types, the host code and kernel code. Host code is executed on CPU and kernel code is executed on GPU. The parallel computation will be contained in the kernel code. Kernel codes can not execute directly by themselves, so the program still starts from the host. And the kernel code will be called by the host when needed, and then the device begins to run the kernel instructions on GPU. After the execution on device is finished, the results will be handled by the host, so the program execution ends on the host.

Each GPU chip consists of several streaming multiprocessors (SM) [8] [10]. The GTX 260 carries 24 individual SMs and the GeForce 8800GTS carries 12 individual SMs. Several streaming processors (SP or core) are contained in each SM: these are the real execution unit. In implementation, to manage hundreds of threads, the SM employs an architecture called SIMT (single-instruction, multiple-thread). The SM maps each thread to one core, and each thread executes independently. The SIMT unit creates, manages, schedules and executes threads in groups of 32 parallel threads called warps [2]. For the current generations GPUs one warp consists of 32 threads.

Fig. 1 illustrates the execution grid model of CUDA. Threads are the atomic unit running on the GPU. Each thread will execute the same kernel code for each different data element. It is similar as the SPMD (Single Program Multiple Data) [11]. Threads are organized into blocks and blocks are organized into grid. Both the block and thread have a maximum dimension of three. Grids and blocks have their size defined by three independent dimensions (x, y and z). For example, in the compute capability version 1.3, a thread block has the maximum dimensions sizes of 512, 512 and 62. A grid has the maximum size of 65535 for each dimension. Because the maximum number of threads in one block is limited to 512, the product of the three sizes of dimensions of the thread block can not be over 512 [8] [12] [13]. So the point of tiling is to find a suitable set of sizes in each dimension which can offer the best performance.

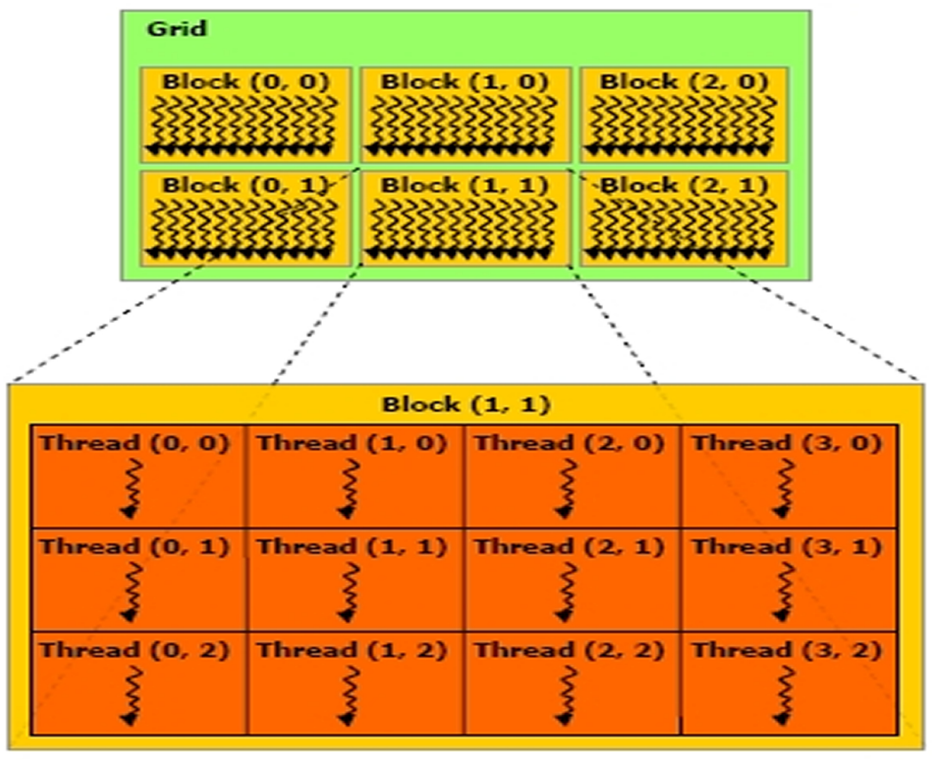

Figure 1. Programming model in CUDA

### B. *Interpolation Algorithm for Image Resizing*

Interpolation algorithms are widely used in many fields, not only for digital image processing. There are several kinds of categories of the interpolation for their implementation such as nearest-neighbor interpolation, bilinear interpolation, bicubic interpolation, fractal interpolation [14] and so on [15]. They use different methods for implementation and have different output quality and efficiency [16].

In this work, we use bilinear interpolation [17] [18] [19] as a sample. The theory of bilinear interpolation algorithm is to use four neighbor pixels to calculate the logical value of a terminal pixel. If $(x_f, y_f)$ are the coordinates of the terminal pixel in final image, $(x_p, y_p)$ are the coordinates of the logical pixel P in source image, $(x_1, y_1)$, $(x_2, y_2)$, $(x_3, y_3)$and $(x_4, y_4)$ are coordinates of four neighboring pixels. The relationship between coordinates $(x_f, y_f)$ and $(x_p, y_p)$ is expressed in (1). The relationship of coordinates between the four neighbor pixels and the logical pixel is expressed in (2) and (3). The offsetY is the offset on the Y-axes direction and the offsetX is the offset on the X-axes direction. Those can be calculated by (4). Then the color information of the logical pixel P can be calculated by (5) and this is the color information for the terminal pixel in final image.

$$x_p = \frac{x_f}{scale}, y_p = \frac{y_f}{scale} \quad (1)$$

$$x_1 = x_3 = \text{int}(x_p), \ x_2 = x_4 = \text{int}(x_p) + 1 \quad (2)$$

$$y_1 = y_2 = \text{int}(y_p), y_3 = y_4 = \text{int}(y_p) + 1 \quad (3)$$

$$offsetX = x_p - x_1, offsetY = y_p - y_1 \quad (4)$$

$$\begin{aligned} f(x_p, y_p) = \\ (1 - offsetY) \times (offsetX \times f(x_2, y_2) + (1 - offsetX) \times f(x_1, y_1)) \\ + (offsetY) \times (offsetX \times f(x_4, y_4) + (1 - offsetY) \times f(x_3, y_3)) \end{aligned} \quad (5)$$

## III. METHOD AND PLATFORM

### A. *Method of Tiling*

There are, in fact, two kinds of tiling techniques, block level tiling and the deeper thread level tiling [6]. Both of these are the logical partition in order to organize the threads according to the data elements. In our experiment we adopt the block level tiling technique for performance tuning.

If each block contains 8x8 threads, each thread calculates one pixel in the final image. The logical partitions should be like Fig. 2.

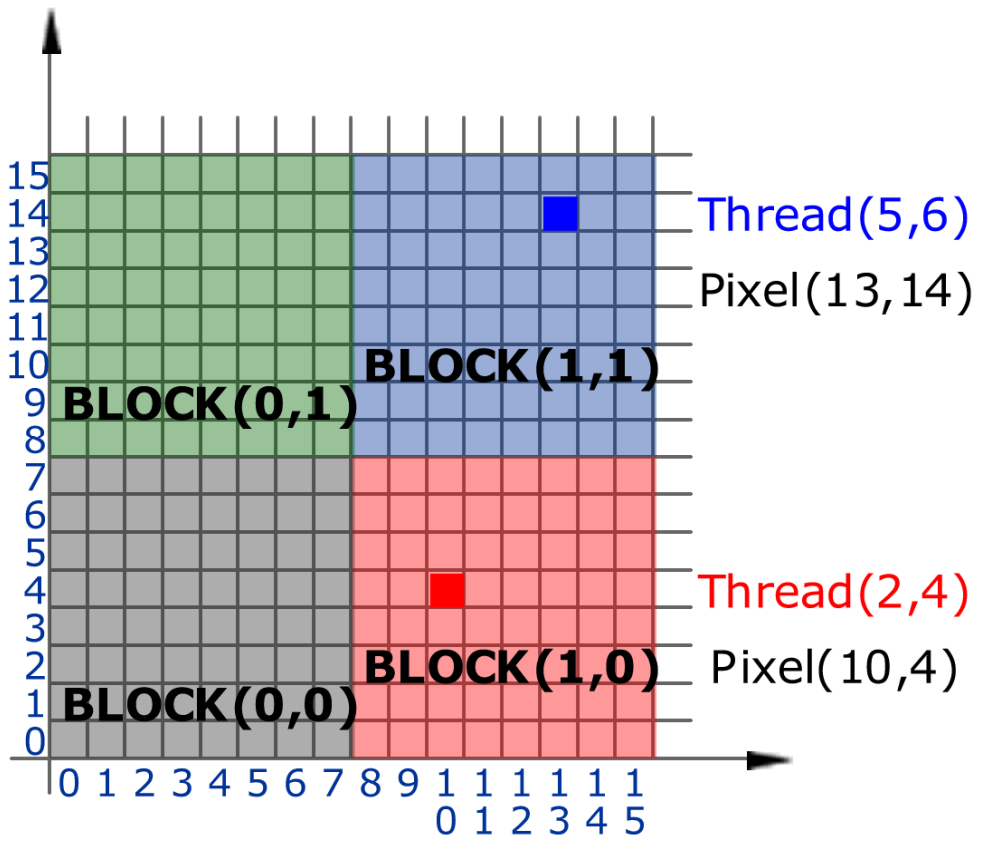

Figure 2. Mapping between threads and data elements

The pixel with coordinates (10, 4) in the final image will be calculated by the thread with thread id (2, 4) in the block with block id (1, 0). The mapping between the threads and pixels can be described by (6).

$$p_x = b_x \times b_{width} + t_x \;,\; p_y = b_y \times b_{height} + t_y \qquad (6)$$

$P_x$ and $P_y$ are the coordinates of a pixel in final image, $b_x$ and $b_y$ are the ids of the block, *b_width* is the width of blocks and the *b_height* is the height of blocks. In Fig. 2 the width and height are both 8, the $t_x$ and $t_y$ are ids of the thread.

### *B. Method of Comparison between GPUs*

Table I lists some differences between the NVidia GTX 260 and the NVidia GeForce 8800 GTS. Assume the scenario when the programmer optimizes the algorithm on the GTX 260; he perhaps sets the tiling dimensions as 32x16 in one thread block. In this case, each SM can have the maximum number of active threads of 1024 within 2 blocks. But this tiling dimension may cause bad performance if running on the GeForce 8800 GTS, because each SM in the GeForce 8800 can only provide a maximum number of active threads per SM of 768. This is not large enough for putting two blocks into one SM, and only one block which includes 512 threads can be placed into each SM. So it can not offer a high utilization and the performance is not as ideal as on GTX 260. And on the other hand, the different features such as number of cores may affect the effectiveness of tiling.

In order to validate it, we execute the same program separately on GTX 260 and GeForce 8800 GTS for a same source image with the size of 800x800 pixels but with different tiling dimensions and different scales. For example, we can execute the program on the first platform with GTX 260 equipped for different tiling strategies, and we can find the tiling dimension TD1 which can provide the best performance. Then the program will be executed on the second platform with GeForce 8800 GTS equipped, and we can also find the tiling dimension TD2 which can provide the best performance. By making a comparison between TD1 and TD2 to be aware of whether they are the same, and what the results are when performing with different scales.

### *C. Developing Platform*

- The hardware environment of the developing platform:

  CPU: Intel(R) Core (TM) 2 Quad 2.83GHz

  System Memory: 3GB

  GPU chip: GTX 260

  GPU Features: see Table 1 and Table 2

- The software environment of the developing platform:

  Operating System: Windows XP Professional (32-bit)

  Development Tools: Microsoft Visual Studio 2005

  Development Language: C++

  CUDA Version: 2.1

### *D. Testing Platform*

- The hardware environment of the first testing platform:

  CPU: Intel (R) Core (TM) 2 Quad 2.4GHz

  System Memory: 4GB

  GPU Chip: GeForce 8800 GTS

  GPU Features: see Table 1 and Table 2

- The software environment of the first testing platform:

  Operation System: Windows XP Professional (32-bit)

  CUDA Version: 2.1

The first development platform is also be used as the second testing platform.

The data needed to be collected is the time spent on the execution of the program which ruing on the GPU. Each item data is the average of 1000 times execution those are divided into 10 groups. Because the offsets on the average to maximum number and minimum number are too small, we don't present error bars on figures.

## IV. TILING FOR PERFORMANCE TUNING

### *A. Tiling on Different GPUs*

As we discussed above, different models of GPUs may have different compute capability versions and features, and these provide different parameters in CUDA. It is absolutely clear that, the GTX 260 can provide better performance than the GeForce 8800 GTS (it is decided by the hardware). But the point here is to find out whether the different features will affect the suitable tiling dimensions which can provide the best performance. Fig. 3 gives the comparisons between GTX 260 and GeForce 8800 GTS.

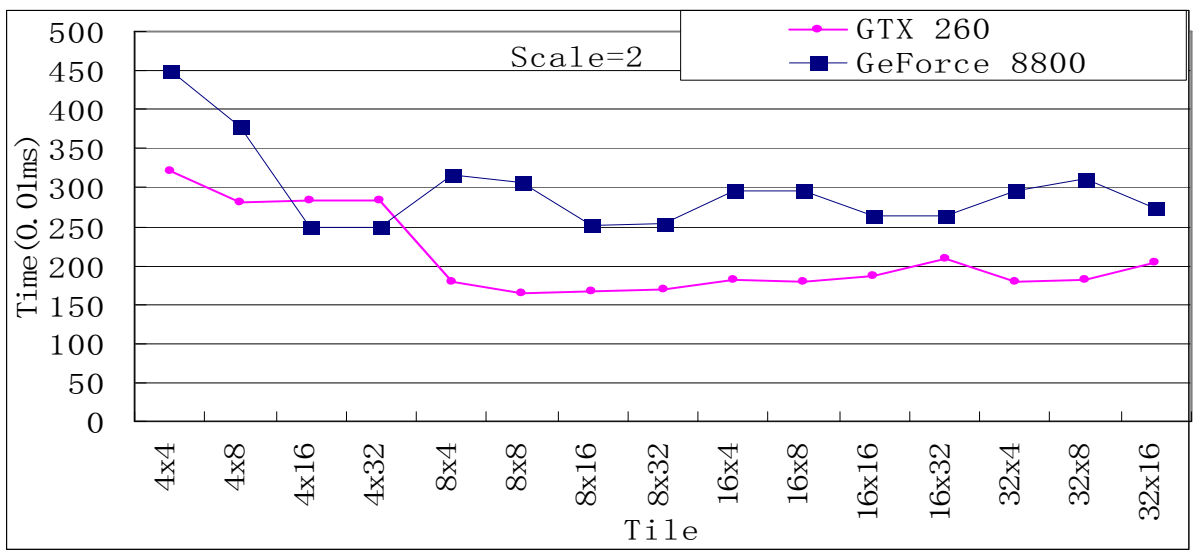

(a)

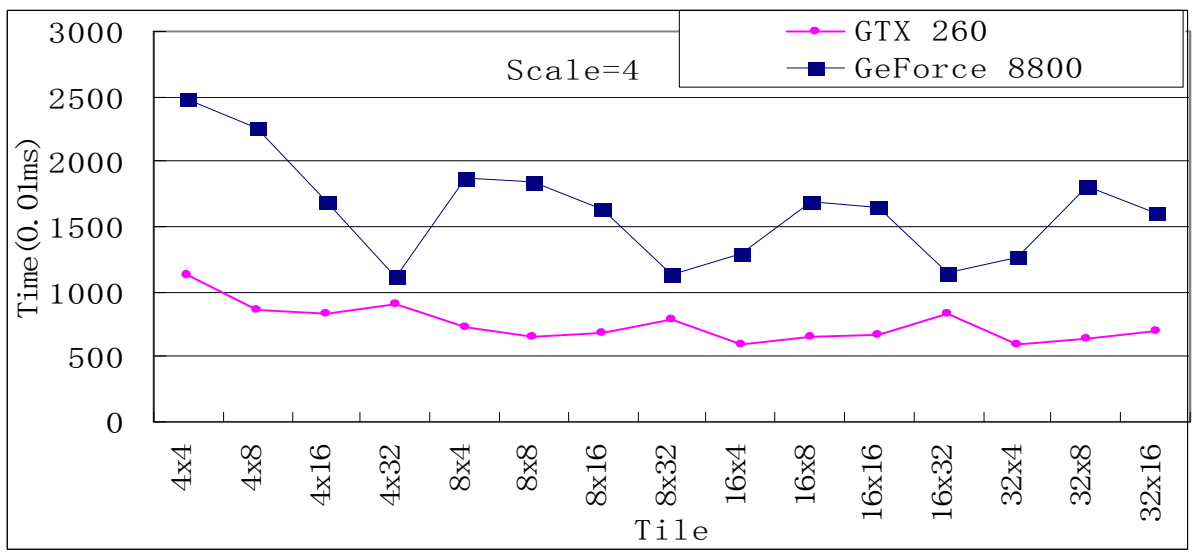

(b)

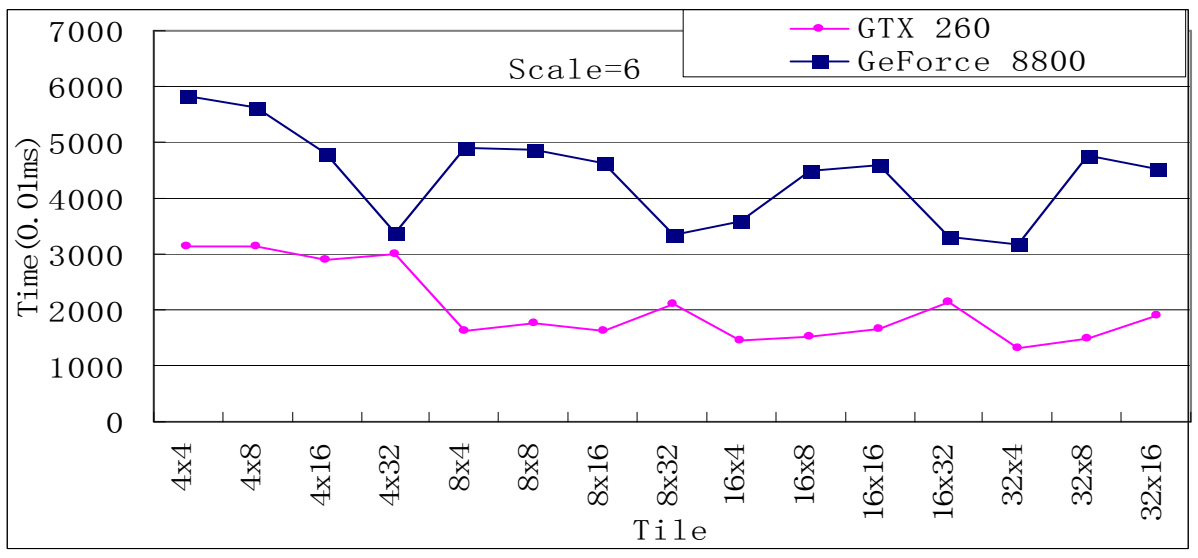

(c)

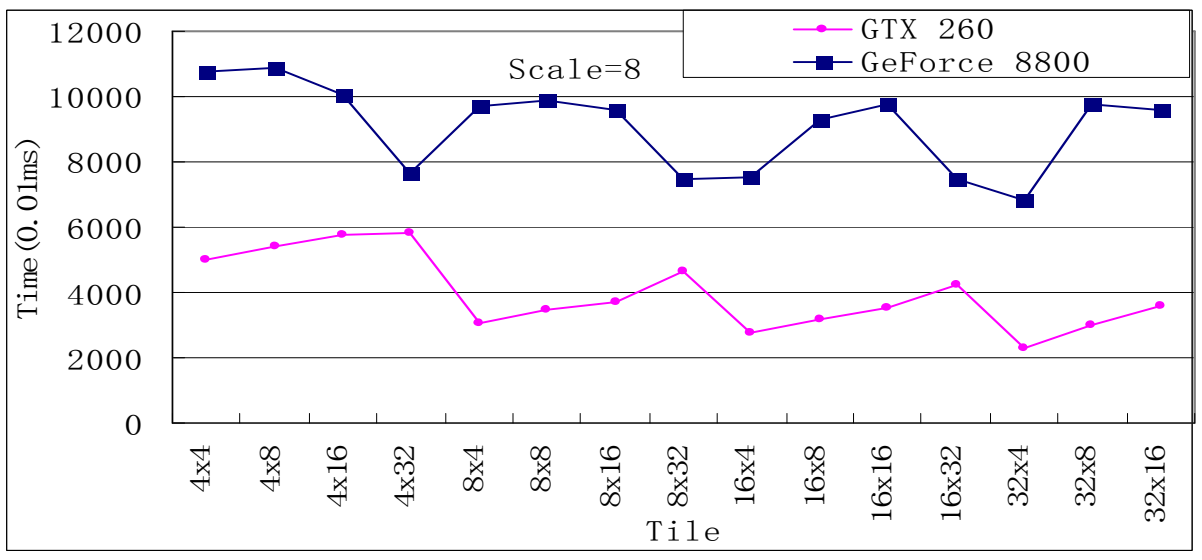

(d)

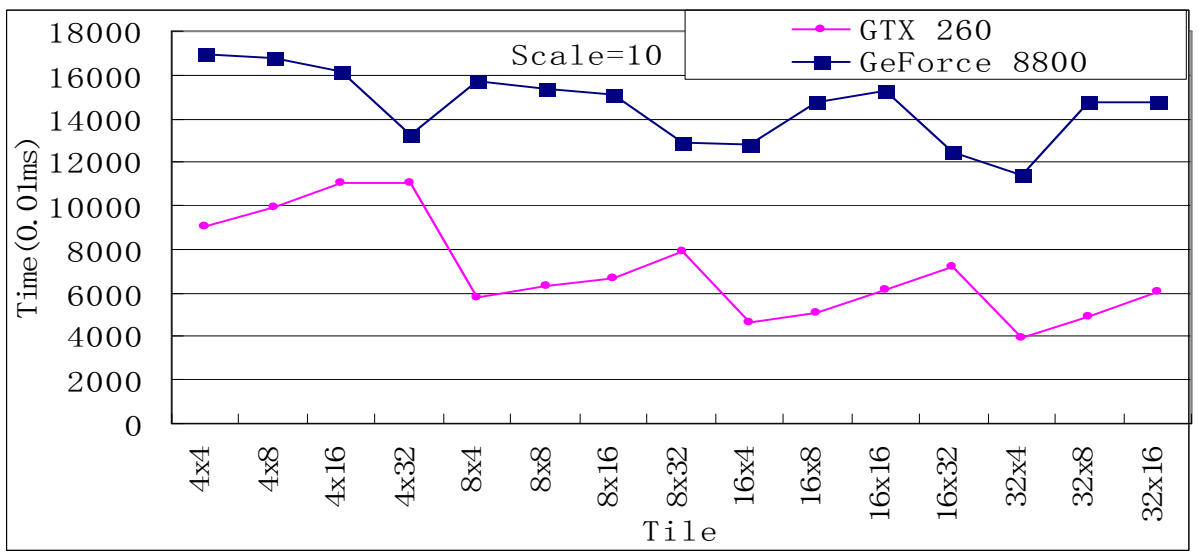

(e)

Figure 3. Comparison between the GTX 260 and the GeForce 8800 GTS for different tiles. The inset (a) to (e) present the results on different scales of 2, 4, 6, 8 and 10

### B. *Reasons for the Effects on Different Platforms*

From Fig. 2, the tiling dimensions which can provide the best performance both on GTX 260 and GeForce 8800 GTX that we can find is the tiling dimensions 32x4 in inset (c), (d) and (e). But in fact, the reason is the growing width of the final image in the last three insets with scales 6, 8 and 10. That means the time spent on pointer moving cross rows occupies more rate of the latency. Since the tiling dimensions 32x4 can provide enough active warps and less vertical memory accessing, so it can bring best performance on both two GPUs. The detailed reason can be explained with reference to Fig. 4.

The two blocks in Fig. 4 both have the same number of threads (32 threads): the left block has the tiling dimensions as 4x8, and the right one has the tiling dimensions as 8x4. During execution, the left solution spends more time because it spends more time on memory accessing. It moves the pointer from one row to next row for each 4; continue accessing and the movement between rows will spend much more time than the movement between columns. But in the right solution, it reduces the movement between rows to 4 times. So the 4x8 tiling dimensions can provide better performance than the tiling dimensions as 8x4. In fact, if the scale is not large then the final image's width is not very large, and the time spent on the movement from one row to another row will not be long. In this case the effect caused by the vertical accessing is not as obvious as in larger final images [20].

So when the vertical memory accessing is not a bottleneck, such as the insets (a) and (b), it presents different appearances. From the five insets, we can find a rough trend that in this situation (we don't have the deep effect of the cost on vertical memory accessing), as shown in insets (a), (b) and (c). The lower line is smoother than the upper line. This means the block size doesn't affect the performance on GTX 260 as significantly as on GeForce 8800 GTS. The reason can be explained as that, the GTX 260 has 192 processors while the GeForce 8800 GTS has the half that number, 96 processors. Compared to GTX 260 the number of parallel threads is a bottleneck on GeForce 8800 GTX: the effect caused by the block size is more obvious than the effect caused by vertical memory accessing. But when the final image is large enough, as shown in insets (d) and (e), the cost on vertical memory will be predominant and this lower line is as jagged as the upper one.

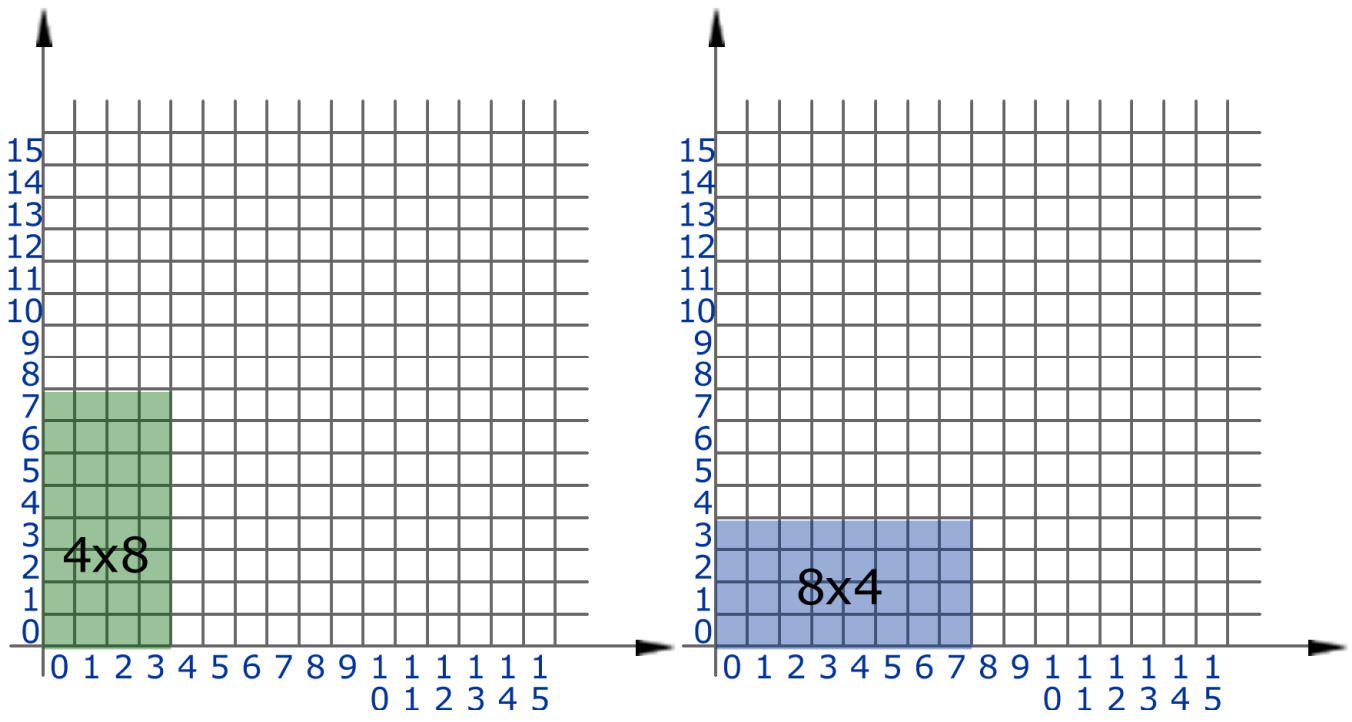

Figure 4. The same size blocks with different tiling dimensions

### C. How Tiling Affect the Performance on Different GPUs

So, the tiling dimensions' effectiveness is also affected by the different compute capabilities. For the bilinear interpolation algorithm, the tiling dimensions of 32x4 can offer better performance when performing on different GPU platforms especially for large scales. But it also can not stand for general situations. Nevertheless, we can have a principle that the more cores the less dependence on tiling dimensions. In order to explain this point more clearly, we give an extreme example as below:

Support G1 is a GPU with two SMs (16 cores), G2 is a GPU with twenty SMs (160 cores). Each SM can support at most 1024 active threads. If one tiling dimensions t2 leads to the half efficiency on each SM. For the G1, because there are two parallel SMs, so every two SMs will lose half efficiency, it will lose 1/4 efficiency totally. But for the G2, because there are twenty parallel SMs, so every twenty SMs will lose half efficiency, it will lose 1/40 efficiency. It is obvious that the effect caused by tiling dimensions is less when the number of cores is larger.

## V. Conclusions

One set of tiling dimensions may not have the same effectiveness when performing on different models of GPUs with different compute capabilities. This is also a new factor that we should consider, and we think it may be a good approach to consider more about the performance on the worst-case GPU in order to let the program get better performance on most GPUs, because the changing on the tiling dimensions brings more effects when performing on the GPU with less cores.

In practice, we think the tiling must be weighed most carefully in CUDA programs' implementation, optimization and compatibility (always get better performance when performing on different GPUs) especially used together with other CUDA techniques. In our experiment, the tiling dimensions 32x4 seems to be a better choice which can offer better performance in general when performing in different situations, especially for larger final images.